# RESERVOIR LAYERS IN HIGH $T_c$ MERCURY CUPRATES[a]

T.H. GEBALLE*, BORIS Y. MOYZHES**, P. H. DICKINSON***
*Dept. of Applied Physics, Stanford University, Stanford, CA 94305, Geballe@stanford.edu
**E. L. Ginzton Laboratory, Stanford University, Stanford, CA 94305
***Chromatic Research Inc., Sunnyvale, CA 94089

## ABSTRACT

We put forward the hypothesis that cations with 6s electrons (Hg,Tl,Pb,Bi) in the charge reservoir layers of high $T_c$ cuprate superconductors actively participate in the pairing interaction as negative-U centers. We further argue that the Hg-cuprates are outstanding superconductors ($T_c > 160$ K) because they can exist as *two-site* negative-U centers, $Hg_2^{+2}$. Their electrons are less localized than in single-site centers (negative-U or bipolaron) and can have a strong pairing interaction with a smaller increase in effective mass. The $Hg_2^{+2}$ centers are oriented in the x and y directions and can have phase differences compatible with the d-wave symmetry of the $CuO_2$ planes.

## INTRODUCTION

In this paper, we examine the occurrence and magnitude of superconductivity in the high-temperature superconducting cuprates. We draw some new conclusions based upon well-established empirical facts, and pay particular attention to the fact that the highest $T_c$s occur when the dopants are in special charge reservoir layers.

Charge reservoir layers are generally favorable for superconductivity because dopants are separated from the $CuO_2$ layers by an intervening nonconducting AO layer (where A is usually Sr or Ba). Random potential fluctuations associated with the doping are better screened and the translational symmetry in the $CuO_2$ planes is better preserved than when the doping is by substitution in layers adjacent to the $CuO_2$ layers, such as in $(LaSr)_2CuO_4$ which has no charge reservoir layer. The spatial separation of the charged dopants from the carriers bears a resemblance to the modulation doping technique employed in high-mobility semiconductor heterostructures. In the following, we argue that the charge reservoir layers play an important role beyond that of doping $CuO_2$ layers.

## EXPERIMENT

### Negative U-Centers In Reservoir Layers

Charge reservoir layers contain heavy nontransition metal cations that act as negative U-centers [1]. We prefer to use *negative-U center* rather than *bipolaron* because negative U-center refers specifically to the situation in which the intermediate valence is unstable with respect to disproportionation. Our hypothesis is that negative-U centers are responsible for the *enhanced* superconductivity observed in cuprates with charge reservoirs. Relevant $T_c$s are given in many places [2]. Empirically, we know that bulk superconductors with the highest $T_c$s have charge reservoirs that contain the metals Hg, Tl, Bi, and Pb, all of which possess 6s electrons. (We exclude from this consideration the infinite layer systems and certain other structures that are not sufficiently well characterized to establish that their superconductivity is a bulk property.) Tl, Bi and Pb have no intermediate valence states between $6s^0$ and $6s^2$. It is well known from their chemistry that paramagnetic ions with $6s^1$ configuration are single site negative-U centers. They disproportionate spontaneously [3] as

$$Me^{+z}(6s^1) + Me^{+z}(6s^1) \rightarrow Me^{z+1}(6s^0) + Me^{z-1}(6s^2), \qquad (1)$$

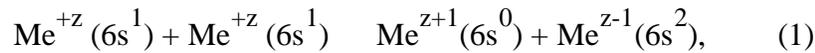

which means that for $Me(6s^0)$ ions there can be an attractive potential for two electrons with opposite spins, and equivalently for the occupied $Me(6s^2)$ ions there can be an attractive potential for holes.

There are two limiting cases of superconductivity associated with negative-U centers. The first occurs when the electrons from the negative-U centers form the band in which the superconductivity occurs. This model is a possible mechanism for the superconductivity in BaKBiO$_3$ and in the tungsten bronzes [4]. It has recently been demonstrated that this model alone cannot account for experimental superconducting data observed in the cuprates [5], although there remains some controversy.

The second type of superconductivity associated with negative-U centers occurs when there are two subsystems of electrons: the itinerant band carriers and the diamagnetic, localized pairs on the negative-U centers. Superconducting pairs increase their binding energy through interaction with the negative-U centers. It is, of course, necessary that, in the superconducting state, the paired carriers on the negative-U centers have the same phases as the itinerant superconducting pairs. Convincing evidence for this second type exists in PbTe doped with Tl [6] for which magnetic and heat capacity experiments demonstrate the existence of two subsystems. One subsystem consists of band electrons and the other consists of the paired electrons on the nearly localized Tl impurities. We examine this evidence in the appendix.

## Mercury Cuprates

Let us now discuss the HgO charge reservoir layers, recalling that $Hg^{+1}$ is different than the other nonvalent $6s^{+1}$ ions. $Hg^{+1}$ exists in condensed phases in the form $Hg_2^{+2}$, a diamagnetic molecular ion [3],

$$Hg^{+1} + Hg^{+1} \rightarrow Hg_2^{+2} \qquad (2)$$

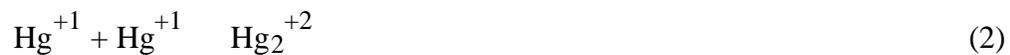

thus forming a two-site negative-U center.

The Hg-containing cuprates with 1, 2, or 3 CuO$_2$ layers per unit cell are commonly designated as Hg-1201, Hg-1212, and Hg-1223, respectively, and have the highest $T_c$s in their respective classes. Furthermore, the pressure dependence of $T_c$ is unique [7] and cannot be explained by calculations of changes in the electronic structure with pressure [8].

The Hg two-site negative-U centers are more favorable for increasing the pairing interactions in the cuprates than the Tl and Bi single-site negative-U centers for two reasons. First, they are oriented along the a- and b-axes. When the a- and b-axis-oriented Hg-Hg centers have opposite phases, they are matched more naturally to the d-wave symmetry of the order parameter than single-site centers. Second, the pairs are less localized in the two-site centers, which results in a smaller effective mass and in more overlap between the $CuO_2$ layers and the HgO layers. The smaller effective mass is important for obtaining high $T_c$ in the Bose Condensation models, in accordance with the Uemera's universal correlation [9]. In any resonant tunneling model the two-site centers are more effective in increasing the coupling between the $CuO_2$ layers.

The role of a possible polaronic effect in the charge reservoir can be tested by searching for an isotope effect. If the lattice is strongly involved then there should be a change in $T_c$ with the mass of the Hg ion.

## **Changes Under Pressure**

From chemistry, it is known that the equilibrium Hg-Hg distance in the $Hg_2^{+2}$ is about 2.5-2.7A [3]. In the reservoir layers, the crystallographic equilibrium distance between Hg ions is 3.9A, but with considerable disorder. For instance, in the Hg-1201 compound, the inplane Debye-Waller factor for Hg is 1.6 sq.$Å^2$ [10], so a sizeable fraction of Hg ions will have nearest neighbors close enough to form strong two-site negative-U centers.

The $T_c$s of the Hg-1201 ($T_c$ = 98 K), Hg1212 ($T_c$ = 126 K) and Hg-1223 ($T_c$ =134 K) all increase with pressure, initially with a linear slope between 1 and 2 K/GPa [7]. They saturate between 20 and 30Gpa, all with a large increase of about 30 K, more than an order of magnitude greater than the corresponding increase for cuprates with Bi and Tl reservoir layers. The large increase in $T_c$ can be understood because the Hg-Hg distance at atmospheric pressure is not optimal, but should become more so under higher pressure, as suggested in Fig. (1). Under pressure the Hg would be expected to shift further from the vertical while preserving the Hg-O(2) distance (as observed), while the unusually large decrease in the Cu-O(2) apical oxygen distance [10] should facilitate additional overlap.

Thus, the two-site negative-U center model gives a simple explanation for the uniquely large increases of $T_c$ under hydrostatic pressure. It also explains why the attempts to use chemical substitution to simulate the effect of pressure have been unsuccessful, even though it works in many other cases [11]. Chmaissem et al have prepared Re,Sr-substituted samples, $(Hg_{0.75}Re_{0.25})Sr_2Ca_2Cu_3O_{8+}$ , which have almost the same interatomic distances as the undoped 1223 under 8.5Gpa [12], and thus might be expected to cause the same behavior as the applied pressure. However, the chemical substitution decreases $T_c$ more than 15 K (from 135 K) versus a rise of 10 K for the pressurized undoped sample. This difference is easily understood in terms of our model because Re brings in extra oxygen, which fills all the vacant sites in the reservoir layer [12], thus precluding the possibility of Hg-Hg two-site negative-U centers.

Neutron scattering investigations have determined the atomic positions of the Hg-containing cuprates, including their pressure dependence. Figure 1 shows the interatomic distances at atmospheric pressure for an optimally doped Hg-1201 sample taken from Balagurov et al [10]. The considerable disorder in the Hg reservoir plane makes it impossible to locate Hg in the plane with any precision. We have shown it to be off-center because of the exceptionally large in-plane Debye-Waller factor, and also to account for the fact that the calculated Hg-O(2) distance of 1.97Å, given in the paper by placing Hg at its crystallographic mean position, is significantly less than the sum of the Hg and O ionic radii of 2.04Å. The difference can be reconciled if the Hg is moved from its mean position by about 0.5Å, as we have indicated in the figure. This is not in disagreement with the scattering data because the uncertainty in planar location of Hg should only permit a determination of the spacing between the Hg and O(2) *planes* The unusually large distance between the $CuO_2$ plane and the apical O(2) can also be understood because the sideways displacement of the Hg allows room for the O(2) to move closer to the positively charged Hg layer. Under pressure all distances diminish, including the in-plane crystallographic mean Hg-Hg distance, and the unusually large Cu-O(2) distance [10]. Because the oxygen sites in the Hg plane are mostly empty, the reduction of the c-axis will cause the Hg atoms to shift in the plane more than the mean lattice constant change thus causing some to become stronger negative-U centers.

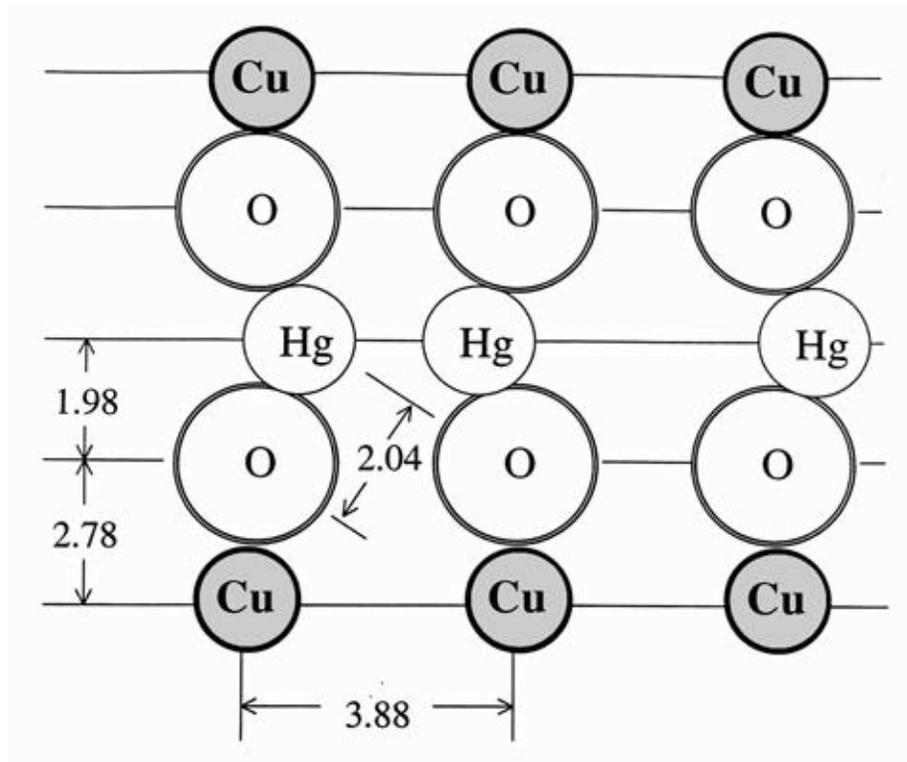

Fig 1.  Schematic cross section of the (100) plane of Hg-1201($Hg_1,Ba_2Cu_1O_{4=}$ ).
The planar distances are for atmospheric pressure [10]. Under pressure the
Cu-O distance is diminished appreciably [10] and we suggest that the
Hg-Hg distances should also change for reasons apparent in the figure.

**CONCLUSION**

Empirical evidence suggests that the Hg, Tl, and Bi-containing electron reservoir layers in the high $T_c$ cuprates contribute to pairing interactions to the superconductivity. Up to the present, most models discussed in the literature, including all bipolaron models, have been devoted to $CuO_2$ layers which are present in all cuprate superconductors. We have argued that complementary pairing interactions in the cuprates and, by inference, in other compounds with complex unit cells, can take place in distinctily different regions (layers). The charge reservoir layers play a role beyond their well-known doping function. The heavy-metal cations in those layers are negative-U centers which have the possibility of phase locking with the $CuO_2$ layers to provide an enhanced pairing interaction and may be responsible for $T_c s > 120$ K. The two-site negative-U centers provided by two properly spaced Hg ions are particularly favorable. They are oriented, as partly shown in Fig. 1, along the x and y directions, and thus are compatible with d-wave symmetry. The application of pressure can further increase $T_c$ 1) because of a more favorable Hg-Hg ion pair separation, and 2) because of an increase in the overlap of the pair wave functions between the reservoir and $CuO_2$ layers. The fact that the Hg layer is so disordered suggests that the superconducting interaction in the Hg-cuprates is still not fully optimized. It might be interesting, for example, to substitute Cd for Hg because $Cd_2^{+2}$ is known to exist with an internuclear distance greater than the $Hg_2^{+2}$ ion [3].

## ACKNOWLEDGEMENTS

We would like to acknowledge a helpful conversation with Professor Seb Doniach. This work was supported in part by the Air Force Office of Scientific Research.

**APPENDIX**

Convincing empirical evidence for negative-U center superconductivity is found in Tl-doped PbTe. PbTe is a 4-6 semiconductor with a direct gap. Tl can substitute on the Pb sites as either a donor ($Tl^{+3}$, $6S^0$), an empty negative-U center, or as an acceptor ($Tl^{+1}$, $6S^2$), a filled negative-U center. The Fermi level is pinned in valence band by the narrow band of the Tl states which are diamagnetic as expected, and which make a dominant contribution to the heat capacity density of states [6].

Superconductivity is observed at temperatures below 1.5 K as a function of Tl doping for concentrations of 0.5% to 1.5% [13]. It is evident that the superconductivity is associated with the Tl states: (i) samples doped over the same concentration range with other donors or acceptors are not superconducting, and (ii) the maximum in $T_c$ occurs at half filling of the Tl states where there are an equal number of filled and empty negative U centers. (The Fermi-level can be swept though the Tl states by additional doping with Na which substitutes on the Pb site as an acceptor.)

The relatively low $T_c$ can be understood in terms of Tl negative U centers and BCS pairing theory as follows: (i) the attraction is localized on the Tl impurities which occupy only a small volume, (ii) the concentration of charge carriers is small, below $10^{20}$ per cc, and (iii) it is necessary to overcome the repulsive interactions between the charge carriers of the host material which prevail throughout the main volume because PbTe is not superconducting when doped with any dopant other than Tl.